\documentclass{PoS}

\usepackage{amsmath}
\usepackage{amssymb}
\usepackage{multirow}
\usepackage{graphicx}
\usepackage{subfigure}

\newcommand{\dd}{\mathrm{d}}

\title{
\vspace*{-2.4cm}
\begin{flushright}\texttt{\footnotesize
BI-TP 2015/16\\}
\end{flushright}
\vfill
A stochastic approach to the reconstruction of spectral functions in lattice QCD}

\ShortTitle{A stochastic approach to the reconstruction of spectral functions in lattice QCD}

\author{\speaker{Hai-Tao Shu}, {Heng-Tong Ding}\\
        Key Laboratory of Quark \& Lepton Physics (MOE) and Institute of Particle Physics,\\
        Central China Normal University, Wuhan 430079, China\\
        E-mail: \email{haitaoshu@mails.ccnu.edu.cn}\\
	E-mail: \email{hengtong.ding@mail.ccnu.edu.cn}}
\author{Olaf Kaczmarek\\ 
	Fakult\"at f\"ur Physik, Universit\"at Bielefeld, 
	33615 Bielefeld, Germany\\
	E-mail: \email{okacz@physik.uni-bielefeld.de}}
\author{Swagato Mukherjee\\ 
	Physics Department, Brookhaven National Laboratory, Upton, NY 11973, USA \\
	E-mail: \email{swagato@bnl.gov}}
\author{Hiroshi Ohno\\ 
	Center for Computational Sciences, University of Tsukuba, Tsukuba, Ibaraki 305-8577, Japan\\
	E-mail: \email{hohno@ccs.tsukuba.ac.jp}}

\abstract{We present a Stochastic Optimization Method (SOM) for the reconstruction of the spectral functions (SPFs) from Euclidean correlation functions. In this approach the SPF is parameterized as a sum of randomly distributed boxes. By varying the width, location and height of the boxes stochastically an optimal SPF can be obtained. Using this approach we reproduce mock SPFs fairly well, which contain sharp resonance peaks, transport peaks and continuum spectra. We also analyzed the charmonium correlators obtained from $N_{\tau}$=96, 48, 32 lattices using SOM and found similar conclusion on the dissociation temperatures of charmonium ground states
 as that obtained using the Maximum Entropy Method.}

\FullConference{The 33rd International Symposium on Lattice Field Theory\\
		14 -18 July 2015\\
		Kobe International Conference Center, Kobe, Japan}

\begin{document}

\section{Introduction} 
The meson SPFs at finite temperature in QCD 
provide us the knowledge to understand the properties of the thermal medium. For example, the thermal dilepton production rate\cite{Braaten90}, which can serve as a (quark-gluon plasma) QGP probe in heavy ion collisions, is related to the SPF in the vector channel directly. The heavy quark diffusion coefficient which describes the propagation of a heavy quark in the medium can also be determined from the slope of the SPF in the zero frequency limit\cite {Kapusta06}. The fate of hadrons can also be read off from the structure of SPF at finite temperature and can serve as a thermometer of QGP\cite{Ding:2015ona,Matsui:1986dk}. The relationship between SPF and Euclidean correlator can be obtained by using
\begin{align}
	\label{eqn_integ_trans}
	G_H(\tau,\vec{p})= \int\limits_0^{\infty}
	\frac{\dd \omega}{2\pi}\rho_H(\omega,\vec{p},T) K(\omega,\tau,T) \quad
	\text{with} \quad
	K(\omega,\tau,T)=\frac{\cosh(\omega( \tau-\frac{1}{2T}))}{\sinh(\frac{\omega}{2T})}.
\end{align}

The extraction of SPF from correlation functions can not be done easily due to the fact that analytic continuation is needed. To describe this function precisely at least $\mathcal{O}(1000)$ points are needed, while the numerical correlator data is 
discrete which contains only $\mathcal{O}(10)$ points. Thus an ordinary {$\chi^2$} fitting is 
inconclusive.~\footnote{However, this approach still works if sufficient information on SPF are known such that number of fitting parameters are much less than that of data points\cite{Ding:2010ga,Francis:2015daa}.} To solve this ill-posed problem several methods have been developed. The most popular method currently used is the Maximum Entropy Method (MEM) which gives the most probable solution based on Bayesian inference\cite{M. Jarrell,Asakawa01}. Other methods, e.g. a new Bayesian approach\cite{Burnier:2013nla}, the Stochastic Analytical Inference (SAI)\cite{H. Ohno} and Backus Gilbert Method (BGM) have been also applied recently\cite{Francis:2015daa,A. Francis}. Since these methods are all based on Bayesian inference, in this paper we will follow an approach not relying on Bayesian inference and introduce a so-called Stochastic Optimization Method (SOM)\cite{A. S. Mishchenko} to extract SPFs from correlators.

\section{Stochastic Optimization Method}
The basic idea of the commonly used MEM is to obtain the most probable SPF from given data by maximizing the conditional probability. A prior information on the SPF requires to be put as a default model in MEM. The most probable SPF obtained from MEM can be proven to be unique, if exists. The general idea of SOM, on the other hand, is to average all possible solutions to obtain the final solution. All these possible solutions are independent with each other and are obtained by minimizing the likelihood function. In this approach, no prior information, such as the "default model" needs to be used. Since the real SPF is positive, the mock SPF initiated needs to be positive.

To perform SOM firstly we define a target function, i.e. likelihood function, in the following form
\begin{align}
	\label{target_function}
	Q=\sum_{i,j=1}^{N_{\tau}}(\overline G_i-\widetilde G_i)[C^{-1}]_{i,j}(\overline G_j-\widetilde G_j) \quad
	\text{with} \quad
        C_{ik}=\sum_{j=1}^{N_{conf}}\frac{(\overline G_i-G_i^{(j)})(\overline G_k-G_k^{(j)})}{N_{conf}\cdot(N_{conf}-1)},
\end{align}
where ${N_{conf}}$ is the number of configurations of the correlator data and $N_{\tau}$ is the number of data points in each configuration. In Eq.(\ref{target_function}) $G^{(j)}$ is the correlator of configuration j and $\overline G$ is the mean value of input correlator. $\widetilde G$ is calculated from one possible solution $\widetilde \rho(\omega,T)$ by the following expression
\begin{align}
\label{simulated_correlator}
\widetilde G(\tau,T)=\int_{0}^{\infty} \frac{\dd \omega}{2\pi} \widetilde \rho(\omega,T) K(\omega,\tau,T).
\end{align}
At a fixed temperature $\widetilde \rho(\omega,T)$ can be parameterized as a sum of many boxes 
\begin{align}
	\label{simulated_spf}
\widetilde \rho(\omega)=\sum_{t=1}^{K} \eta_{\{P_t\}}(\omega)
\quad	\text{with} \quad
\eta_{\{P_t\}}(\omega)=\begin{cases}h_t, \omega \in [c_t-w_t/2,c_t+w_t/2]\\0, \text{ otherwise} \end{cases}
\end{align}
where $w_t$, $h_t$, $c_t$ are width, height and center of a box. If two boxes overlap, the heights of the two boxes should be added up in the overlapped region. 
 
For each attempt to obtain one possible solution $\widetilde \rho (\omega)$ we focus on two aspects. The first is the generation of different configurations of boxes. This can be realized by eight different elementary updates which reshape these boxes. The second is the selection of generated configurations. These operations are performed aiming to minimize the target function Q defined in (c.f. Eq. (\ref{target_function})). To avoid falling into local minima when minimizing Q the Simulated Annealing Algorithm (SAA) is adopted. In SAA the cooling process is controlled by a fictitious temperature which decreases as $T_{i+1}=T_i\cdot decayscale$ and finally stops at a rather small value $T_{stopping}$ close to 0, like 0.0001 in practice. This temperature determines whether to accept or reject the generated configuration with a probability $P_{adopt}=min\{1,exp(-\delta Q/T_i)\}$ where $\delta Q$ is the difference of Q between two successive elementary updates. The number of elementary updates performed for each temperature is called $markovlength$. And it can be proven theoretically that the global minima can be reached if: (I) $T_{stopping}$ is close enough to 0, (II) $decayscale$ is close enough to unity, (III) $markovlength$ is long enough\cite{S. Kirkpatrick}. The configuration given at $T_{stopping}$ is regarded as a possible solution. By repeating the whole procedure L times and then averaging over these L possible solutions an optimal SPF can be obtained.
 
By elementary updates we mean a random change of the parameter sets $\{{P_t}\}=\{h_t,w_t,c_t\}$ of the boxes. During the updates the number of the boxes are kept fixed. And the change of parameters must satisfy the domains of definitions of a box $\Xi$, are ${h_t} \in [h_{min},h_{max}]$, ${w_t} \in [w_{min},w_{max}]$ and ${c_t} \in [\omega_{min},\omega_{max}]$. Each elementary update of the optimization procedure is organized as a proposal to change some continuous parameter $\xi$ by a randomly generated change $\delta\xi$ in a way that the new value $\xi+\delta\xi$ belongs to $\xi$'s domains of the definition $\Xi_\xi$. To minimize the target function more efficiently we use a parabolic interpolation to find an optimal value of the parameter change $\xi_{opt}=-b/2a$, which gives the extremum value of the fitting function $f(\xi)=a\xi^2+b\xi+c$. The coefficients $a,b,c$ can be obtained by fitting to points ($\xi$,Q($\xi$)), ($\xi+\delta\xi$,Q($\xi+\delta\xi$)) and ($\xi+\delta\xi/2$,Q($\xi+\delta\xi/2$)). In the case $a>0$ and $\xi_{opt}\in\Xi_\xi$ we adopt the increment which gives the smallest target function value among $\delta\xi$, $\delta\xi/2$ and $\delta\xi_{opt}$. Otherwise, if $\xi_{opt}$ is outside $\Xi_\xi$ or $a\leq0$, one just need to adopt Min(Q($\xi+\delta\xi$), Q($\xi+\delta\xi/2$)). The elementary updates used in SOM are listed as follows:
 
(I) \emph{Shift box.} Vary the center $c_t$ of a randomly selected box stochastically. The continuous parameter is restricted to be in the domain of definition $\Xi_{ct}=[\omega_{min},\omega_{max}]$. The part of the box which goes beyond $\Xi_{ct}$ is treated to be a new box whose left side located at $\omega_{min}$.

(II) \emph{Lengthen/shorten box.} Increase/decrease the height of a randomly chosen box keeping the center $c_t$ fixed. The continuous parameter $\xi=h_t$ is restricted by $\Xi_{ht}=[h_{min},h_{max}]$. 
 
(III) \emph{Broaden/narrow box.} Increase/decrease the width of a randomly chosen box keeping the center $c_t$ fixed. The continuous parameter $\xi=w_t$ is restricted by $\Xi_{wt}=[w_{min},w_{max}]$.
 
(IV) \emph{Swap part of the height/width between two boxes.} Choose two boxes A and B randomly. Cut part of the height/width of A and add this part to B. The centers of box A and box B are fixed. The continuous parameter $\xi$ is set to be the height/width of the box A, i.e. $h_t(A)$/$w_t(A)$ which is restricted by $\Xi_{ht}=[h_{min},h_{max}]$/$\Xi_{wt}=[w_{min},w_{max}]$. Meanwhile the height/width of the box B $h_t(B)$/$w_t(B)$ must also be restricted by $\Xi_{ht}$/$\Xi_{wt}$. The area can be changed if needed. This update aims to make a connection between two boxes helping to avoid the local minima more efficiently.  
\section{Analysis with mock data}
To check the applicability of SOM to extract SPF and its dependence on the quality of lattice data, we test SOM with mock data first. Unless clearly pointed out, the test results shown below are for a single solution and $N_{\tau}$ is set to be 96. The error on the mock data is chosen to Gaussian-distributed with $error=0.0001\cdot G(\tau)\cdot \tau$.
 
\begin{figure}[h]
	\centering
	\includegraphics[width=0.43\textwidth]{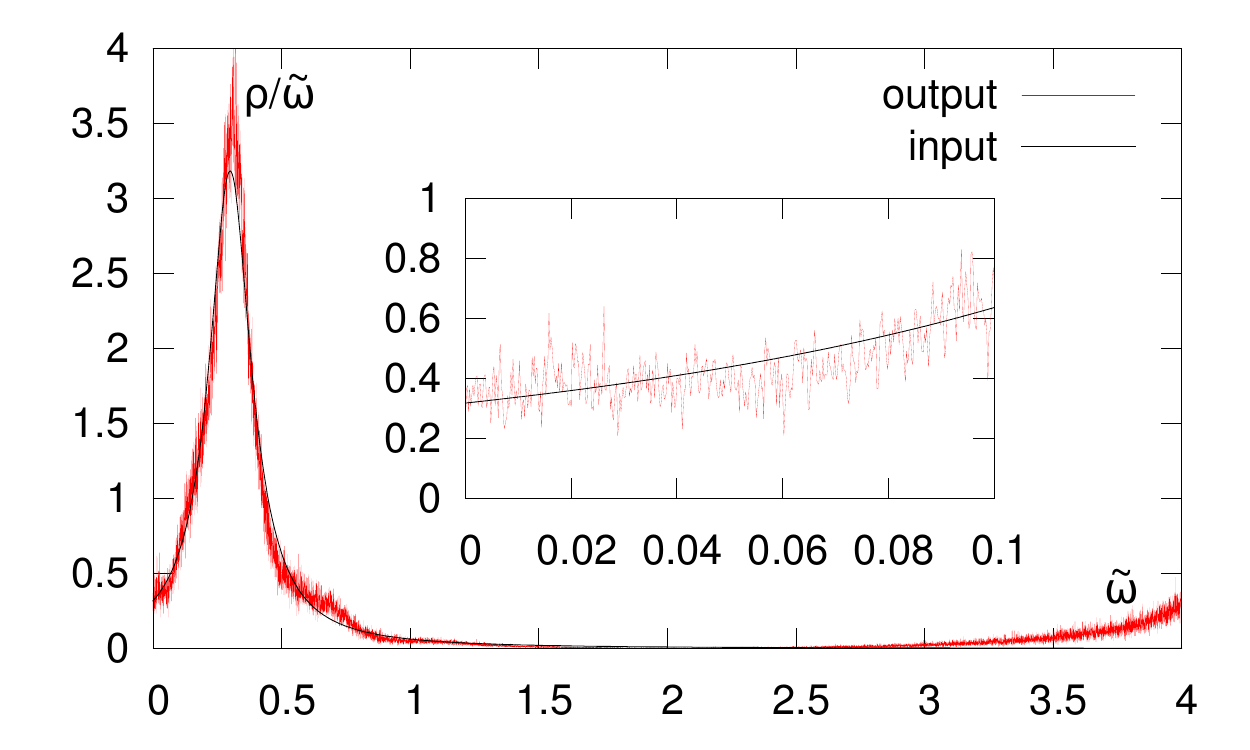}
    \includegraphics[width=0.43\textwidth]{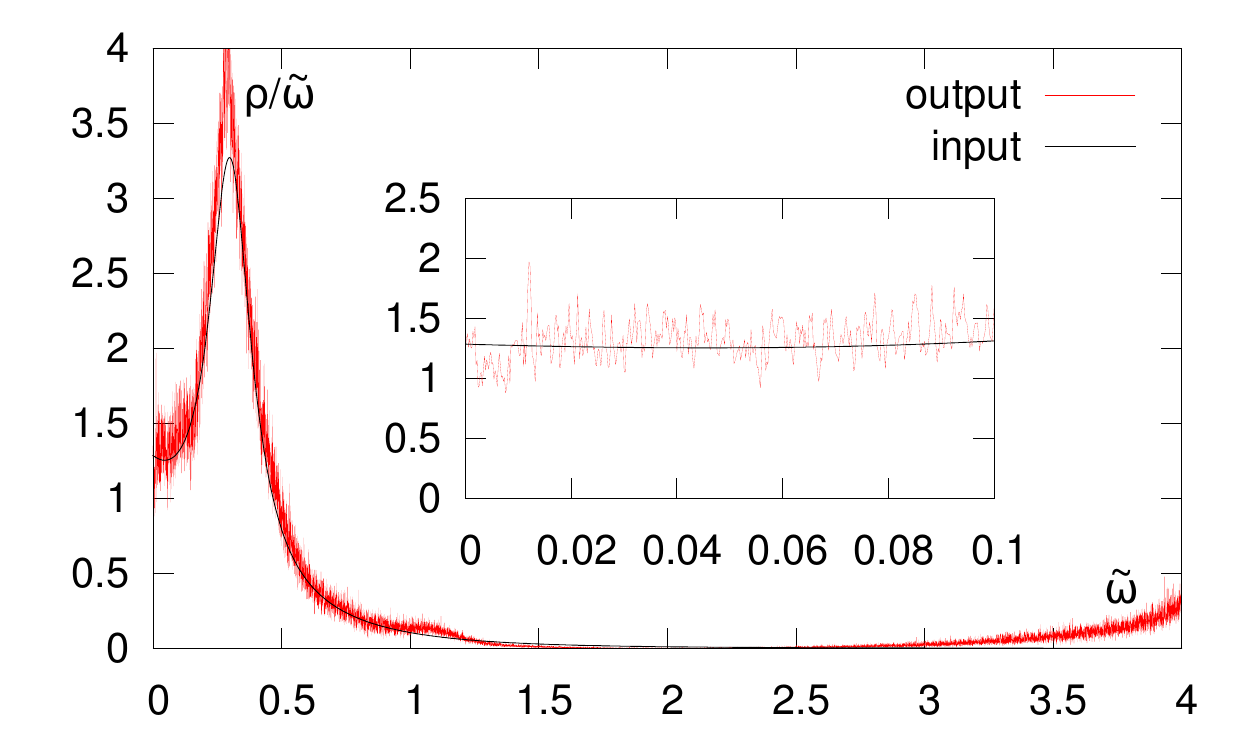}
	\caption{ Mock data test with a BW peak only. \textit{Left}: Intercept is set to be 0.32. The inset shows for a small $\tilde{\omega}$ range to show the intercept clearly.   \textit{Right}: Same as the left plot but intercept is increased to 1.32.}
	\label{intercept}
\end{figure}

The general structure of SPF includes a transport peak, several resonance peaks and a continuum part. The transport peak is reflected in the intercept of a SPF. So first we start from a simple case only with one broad Breit-Wegner (BW) peak in SPF 
\begin{align}
\label{breit-wegner}
\rho(\tilde{\omega})=\frac{\tilde{\omega}\eta}{\pi((\tilde{\omega}-M)^2+\eta^2)}
\end{align}  
with a moderate intercept. Here $M$ is the peak-location of the BW peak. The reproduced SPF is shown in left hand side of Fig.~\ref{intercept}. We can see that the output SPF from SOM can reproduce the input SPF, i.e. the broad BW peak very well. Then we increase the intercept by adding an additional term $a\tilde{\omega}exp(-3\tilde{\omega})$ to the mock SPF to see how SOM performs. This term does nothing but gives a contribution to the intercept since it decreases rapidly in large $\tilde{\omega}$ region. The constant $a$ controls the magnitude of the additional term. As we can see from the right hand side of Fig.~\ref{intercept} SOM still works very well in this case. 

\begin{figure}[h]
	\centering
	\includegraphics[width=0.43\textwidth]{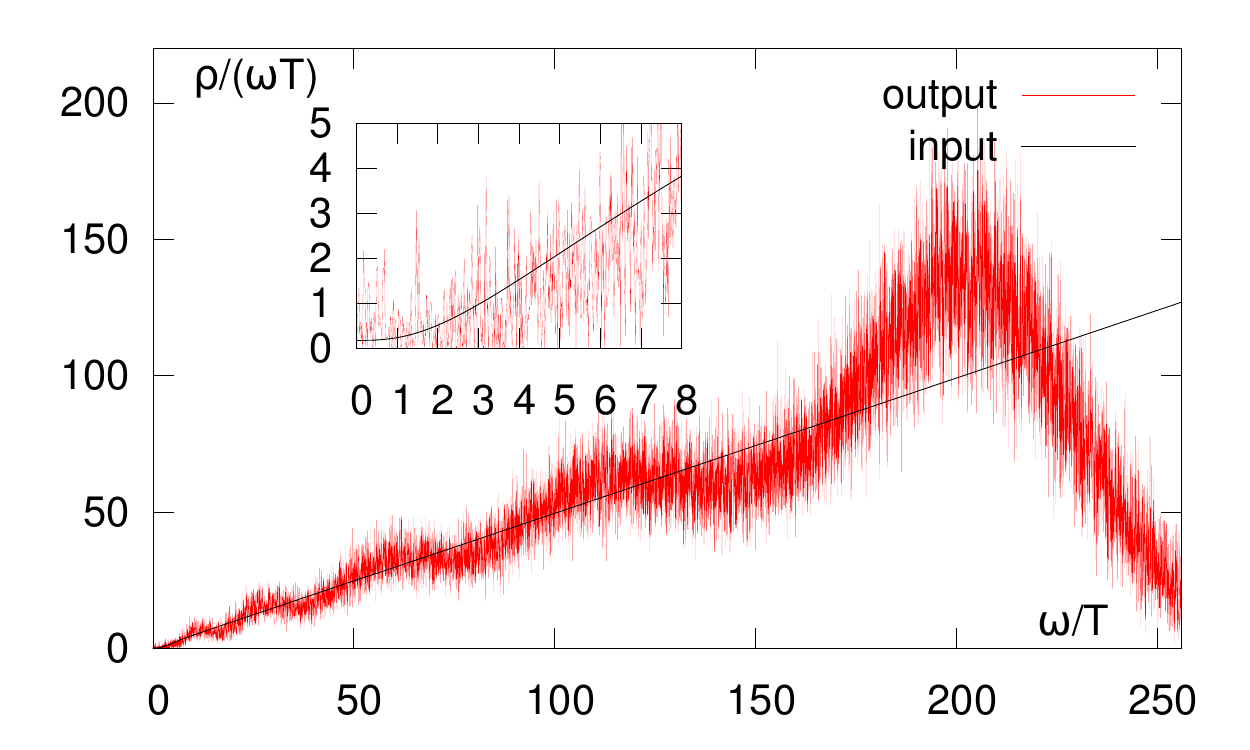}
    \includegraphics[width=0.43\textwidth]{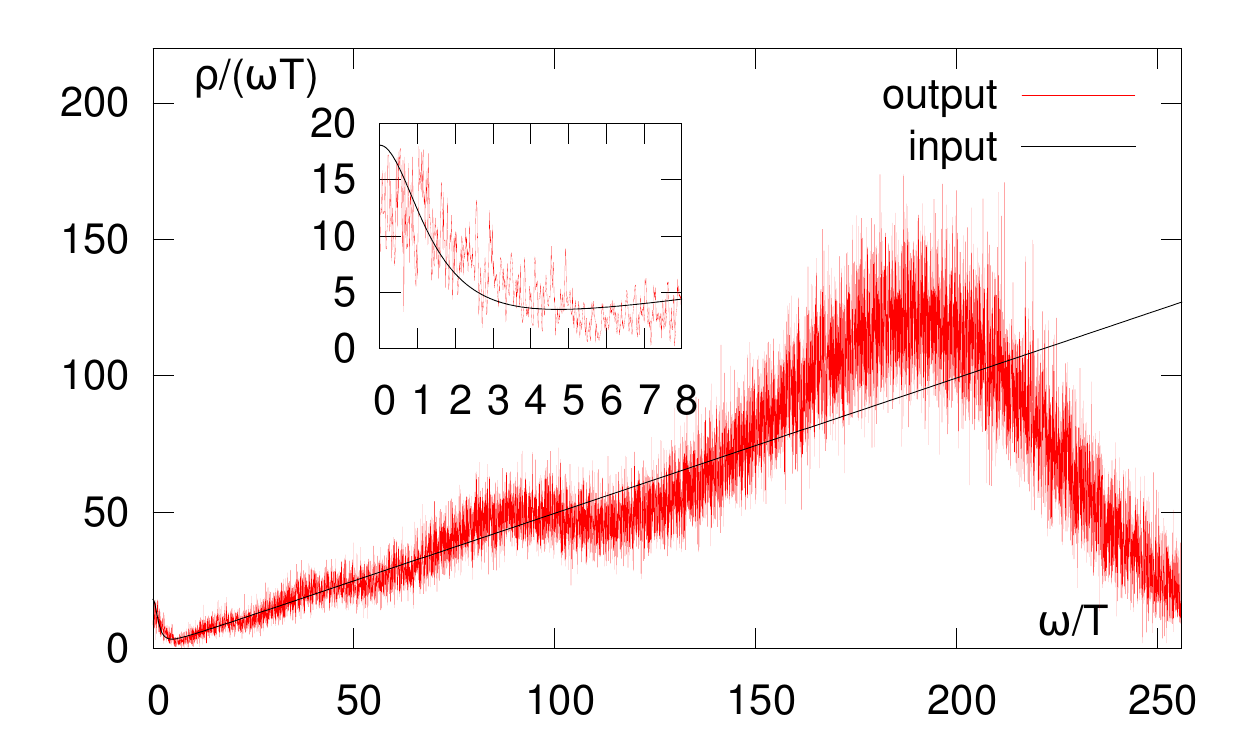}
	\caption{ Mock data test with a BW peak + continuum. \textit{Left}: Intercept is set to be 0.17. The small figure inside is for a small $\tilde{\omega}$ range to show the intercept clearly.   \textit{Right}: Same as the left plot but intercept is increased to 17.}
	\label{florian}
\end{figure}

Next we add an increasing continuum to a transport peak as follows
\begin{align}
\label{increasing_continuum_intercept1.7}
\frac{\rho}{\omega T}=\frac{a\eta }{(\frac{\omega}{ T})^2+\eta^2}+b\cdot (\frac{\omega}{ T})tanh(\frac{4\omega}{ T})
\end{align}
to see the influence of the continuum part on the reconstruction of the intercept. As shown in Fig.~\ref{florian} we can see a too large/small intercept cannot be reproduced very well as the continuum gives the dominant contribution to the correlator. The reason for a large intercept is that a lot of boxes are needed around the boundary, however, for a stochastic method boxes are more inclined to move right to contribute to the large $\omega$ region. For a small intercept, the value of this intercept is comparable to the maximum height of a single box so the noise dominates in the small $\omega$ region. Anyway, in both cases the continuum part is reproduced except for some fluctuations.

\begin{figure}[h]
	\centering
	\includegraphics[width=0.43\textwidth]{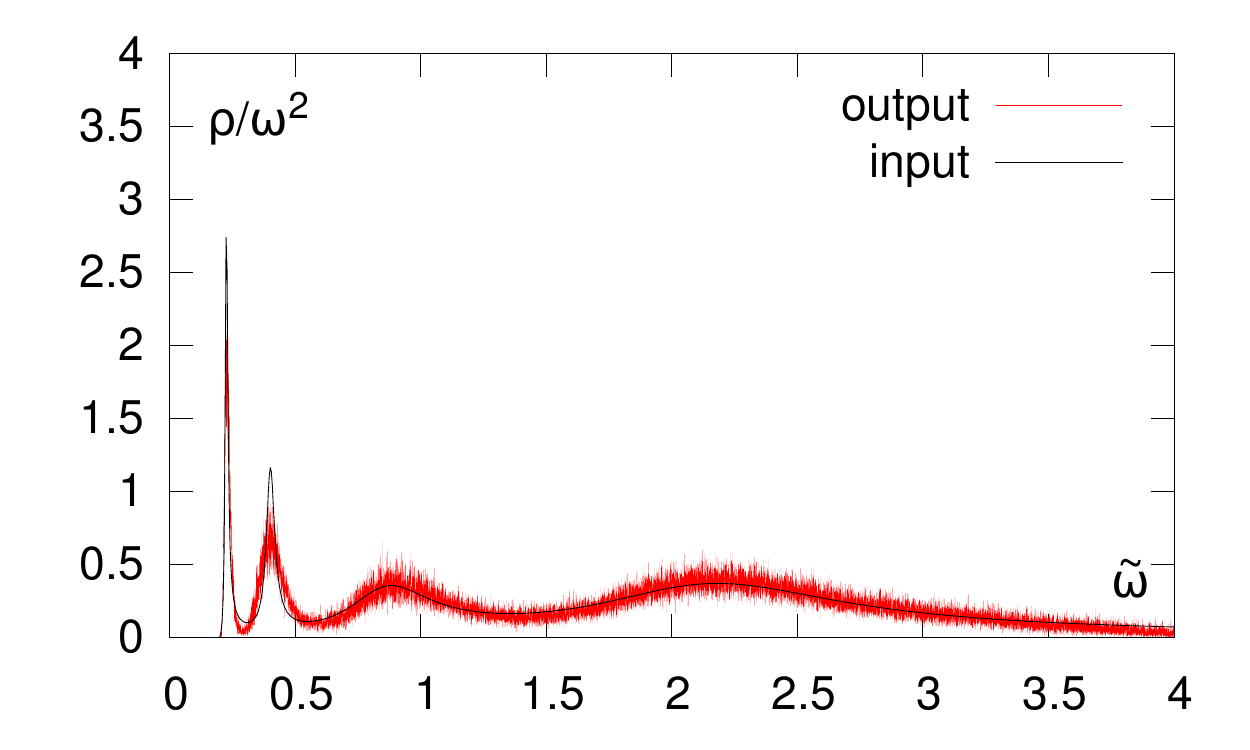}
    \includegraphics[width=0.43\textwidth]{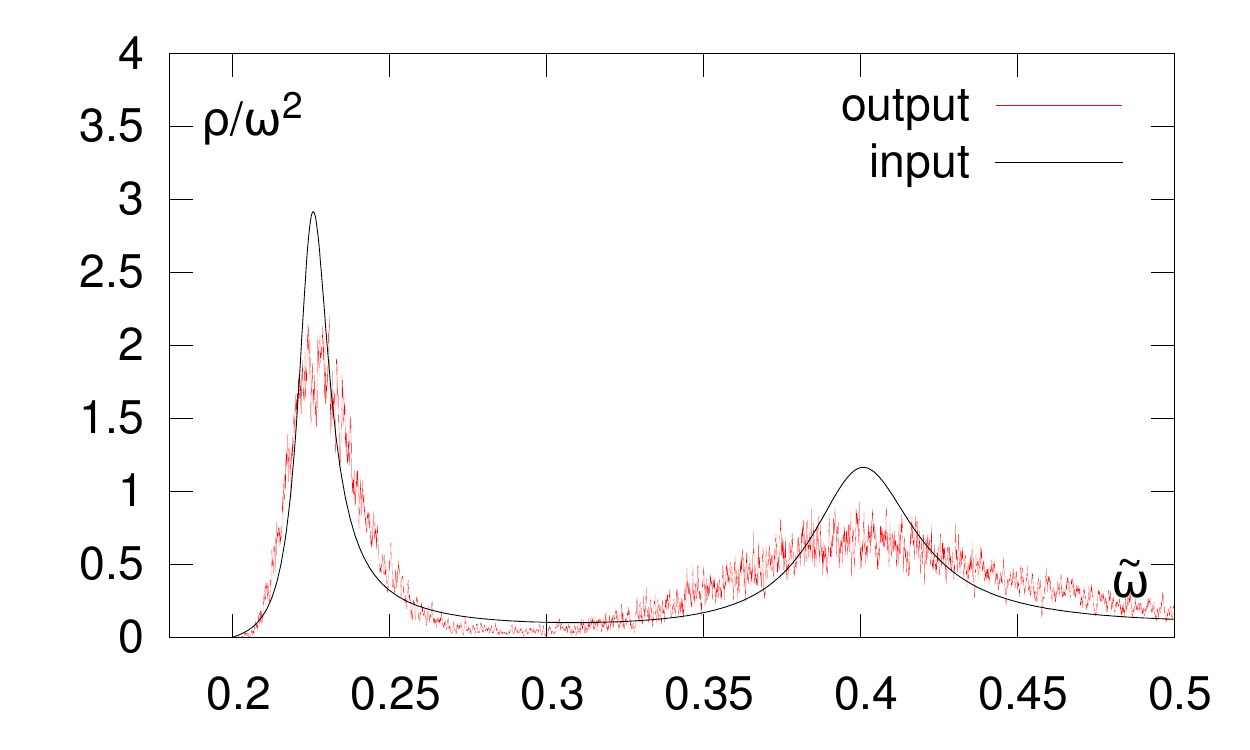}
	\caption{ Mock data test with 4 resonance peaks. \textit{Left}: Whole $\tilde{\omega}$ range.   \textit{Right}: Small $\tilde{\omega}$ range.}
	\label{four_resonances}
\end{figure}

Now we look at the case with more resonance peaks only. This time we have four modified BW resonances and the first two are relatively sharp. To mimic the situation at $T<T_c$ we do not introduce a transport peak. The mock SPF is in the form of
\begin{align}
\label{4resonances}
\frac{\rho}{\omega^2}=\Theta(\tilde{\omega}-\epsilon_{th})\sum_{i=1}^4\frac{a_i\cdot(\tilde{\omega}-\epsilon_{th})}{(\tilde{\omega}-M_i)^2+{\eta_i}^2}
\end{align}  
starting at the threshold $\epsilon_{th}$. Here $a_i$ is constant and $M_i$ denotes the peak-location of the resonance. It can be seen from Fig.~\ref{four_resonances} that although the peak heights of first two resonances are a bit lower than mock SPF the general features of SPF can be reproduced very well by SOM. 

\begin{figure}[h]
	\centering
	\includegraphics[width=0.43\textwidth]{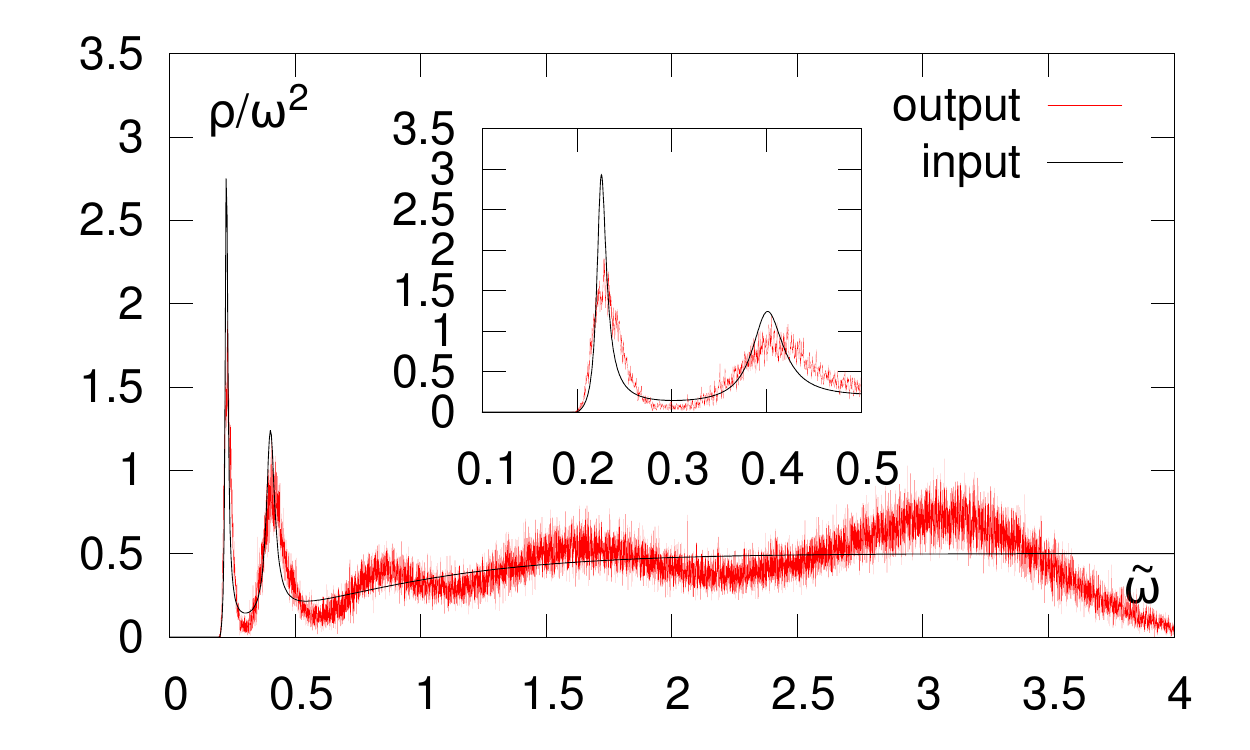}
        \includegraphics[width=0.43\textwidth]{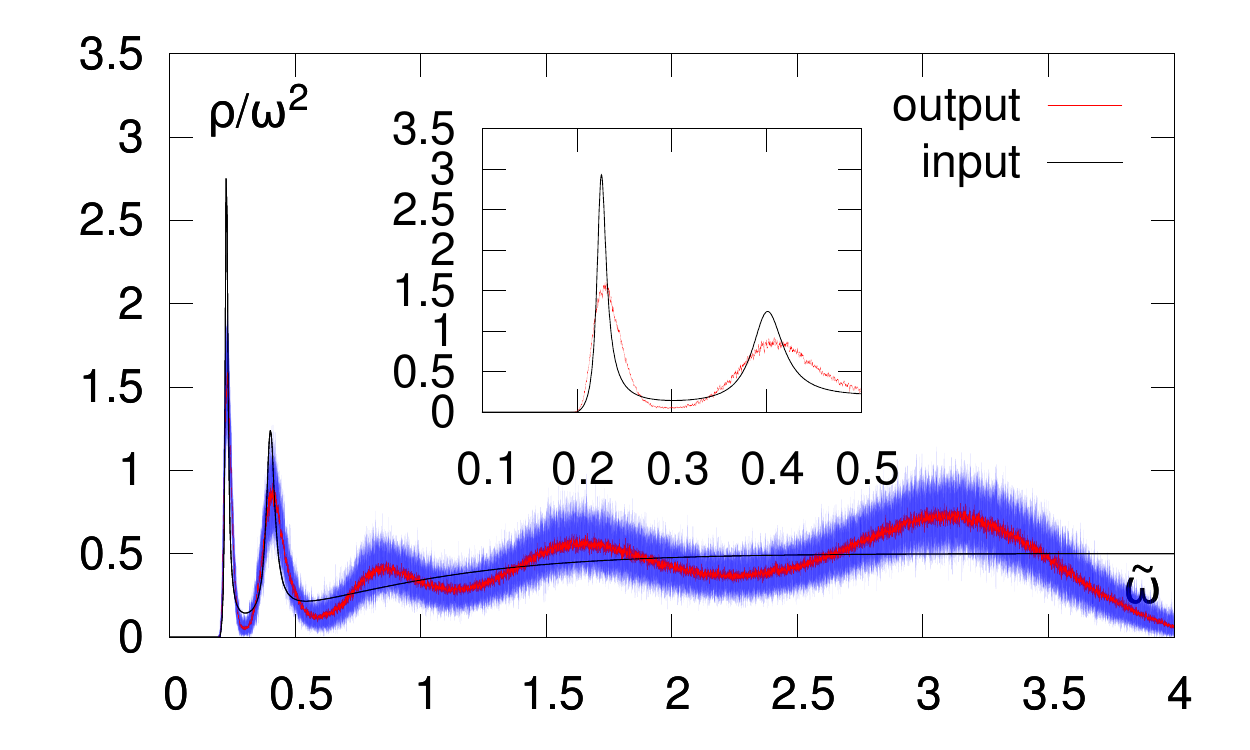}
	\caption{ Mock data test with 2 resonance peaks + continuum. \textit{Left}: A Single solution.   \textit{Right}: The final solution obtained by averaging over 100 solutions.}
	\label{2res_conti}
\end{figure}

At last we consider the influence of continuum part on sharp resonance(s). In this test we use two sharp BW resonances and an almost constant continuum
\begin{align}
\label{2resonances_continuum}
\frac{\rho}{\omega^2}=\Theta(\tilde{\omega}-\epsilon_{th})\bigg(\sum_{i=1}^2\frac{a_i\cdot(\tilde{\omega}-\epsilon_{th})}{(\tilde{\omega}-M_i)^2+{\eta_i}^2}+b\cdot tanh(\tilde{\omega}-\epsilon_{th})\bigg).
\end{align}
As we can see in the left hand side of Fig.~\ref{2res_conti} that the first two resonances are reproduced well and the continuum part also appears except for the fluctuations.
We conclude that the constant continuum will not affect the sharp resonances much.  
To suppress fluctuations we average over 100 solutions and the result is shown in the right hand side of Fig.~\ref{2res_conti}. 

\section{Real lattice data results}

\begin{figure}[h]
	\centering
    \includegraphics[width=0.43\textwidth]{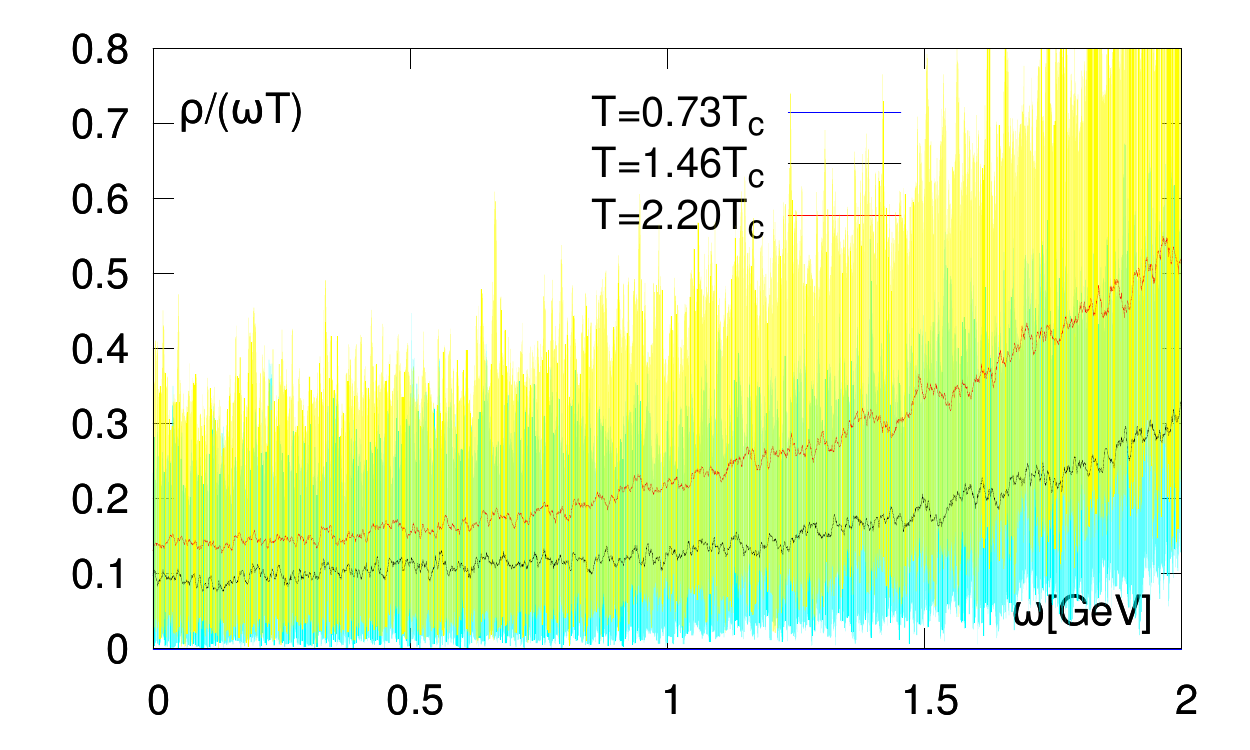}
    \includegraphics[width=0.43\textwidth]{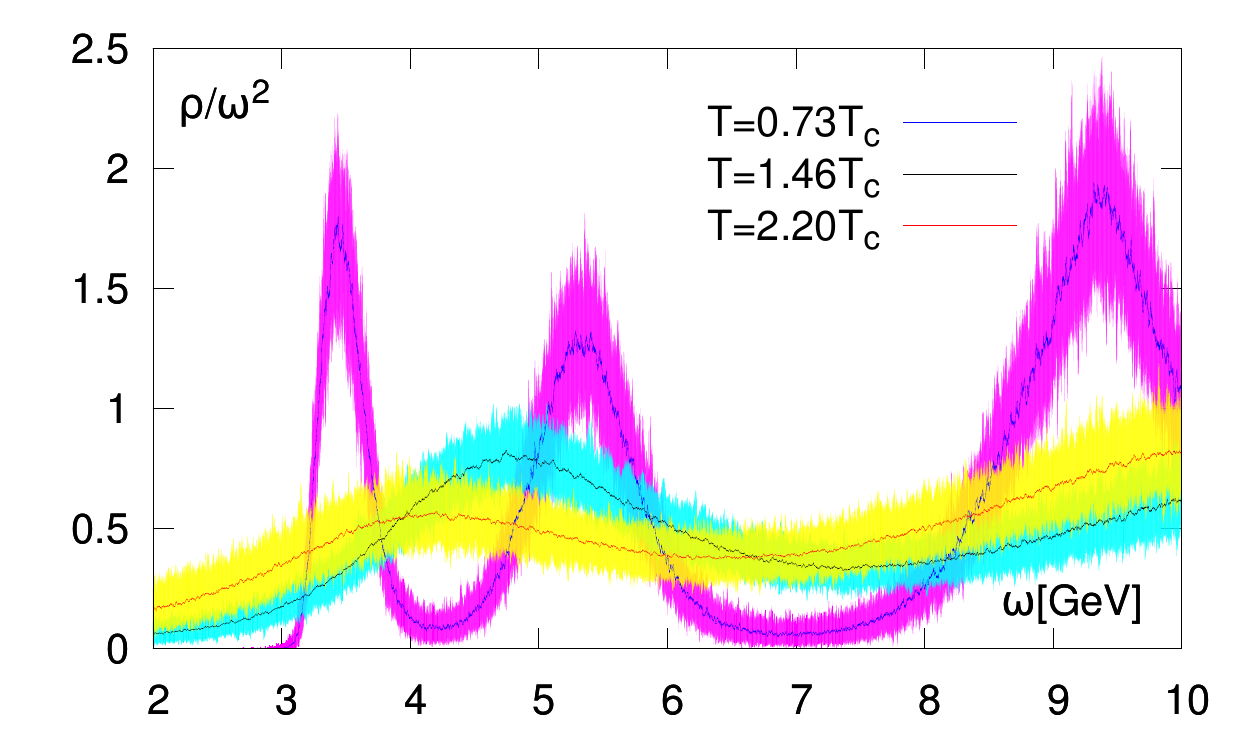}
	\caption{ SPFs of charmonium in vector channel at different temperatures.}
	\label{real_som}
\end{figure}

We applied SOM to the real lattice data taken from Ref.\cite{Ding:2012sp} with $N_{\tau}$=96, 48, 32. Fig.~\ref{real_som} shows the SPFs of charmonium in vector channel at $T=0.73T_c,1.46T_c$ and $2.20T_c$. This result is averaged over 100 possible solutions. As shown in the left plot, the transport peak is obtained but it has to be pointed out that the accuracy of this transport peak can still be improved. From the right plot we know the first peak-location below $T_c$ is 3.44 GeV which is similar to the screening mass extracted from the corresponding spatial correlation function, i.e. 3.472 GeV. It can be also clearly seen that at $T>T_c$ the transport peak is obtained and as temperature increases, the resonance peaks start to disappear. Tab.~\ref{tab_comparison} lists the locations (in [GeV]) and peak-heights of first peaks  of charmonium in the vector channel at different temperatures obtained from SOM and MEM\cite{Ding:2012sp}. We can see that the peak-locations obtained from SOM are quite similar to those from MEM while the peak-heights by SOM are larger. 

\begin{table}[h]
	\centering
\begin{tabular}{|c|c|c|c|}
		\hline
		 & $0.73T_c$ & $1.46T_c$ & $2.20T_c$ \\ \hline
		$SOM$ & $3.44/1.75$ & $4.8/0.72$ & $4.1/0.57$ \\ \hline
		$MEM$ & $3.48/0.98$ & $4.7/0.58$ & $--$\\ \hline
	\end{tabular}
\caption{Peak-locations in [GeV] (the first value) and peak-heights (the second value) from SOM and MEM.}
\label{tab_comparison}
\end{table}

\section{Conclusion}

A Stochastic Optimization Method is proven to be reliable when tested by mock SPFs including transport peaks, resonance peaks and continuum spectra. When applied for extracting spectral functions from temporal Euclidean correlation functions obtained in lattice QCD, SOM also gives results comparable to those from MEM. In the near future we will apply this method to temporal correlators obtained on finer lattices.  

\section{Acknowledgement}
HTS is grateful to members of BNL-Bielefeld-CCNU collaboration for many useful suggestions. SM is supported by U.S. Department of Energy under Contract No. DE-SC0012704.


\end{document}